\documentclass[aip,apl,twocolumn,preprintnumbers,amsmath,amssymb]{revtex4}

\usepackage{graphicx}
\usepackage[colorlinks=true,citecolor=blue]{hyperref}

\begin{document}
    %\linenumbers
    %\title{Interfacing a single quantum dot spin to an optical nano-cavity}
    \title{Coupling an electron spin in a semiconductor quantum dot to an optical nano-cavity}
    \author{Arka Majumdar$^1,^\dag$}
    \email{arka.majumdar@berkeley.edu}
    \author{Per Kaer $^2$}
    \author{Michal Bajcsy$^1$}
    \author{Erik D. Kim$^1$}
    \author {Konstantinos G. Lagoudakis$^1$}
    \author {Armand Rundquist$^1$}
    \author{Jelena Vu\v{c}kovi\'{c}$^1$}
    \affiliation{$^1$E.L.Ginzton Laboratory, Stanford University, Stanford, CA, $94305$\\
    $^2$ DTU Fotonik, Department of Photonics Engineering, Technical University of
Denmark, Building $345$W, $2800$ Kgs. Lyngby, Denmark\\
    $^\dag$ Present address: Physics Department, University of California, Berkeley, CA, $94720$}
    %\date{\today}
\begin{abstract}
We propose a scheme to efficiently couple a single quantum dot
electron spin to an optical nano-cavity, which enables us to
simultaneously benefit from a cavity as an efficient photonic
interface, as well as to perform high fidelity (nearly 100\%) spin
initialization and manipulation achievable in bulk semiconductors.
Moreover, the presence of the cavity speeds up the spin initialization
process beyond GHz.

\end{abstract}
\maketitle
%\section{Introduction}
Single quantum emitters coupled to optical cavities constitute a
platform that allows conversion of quantum states of one physical
system to those of another in an efficient and reversible manner.
Such coupled, cavity quantum electrodynamic (cQED)systems have
been the subject of intense studies as one of the building blocks
for scalable quantum information processing and long distance
quantum communication \cite{Kimble-internet2008, rempe_entangle_2012}. Over the years,
these systems have evolved from conventional atomic systems
\cite{Kimble94} to include those based on solid state emitters, such as
quantum dots \cite{article:yoshio04, andrei_njp}, nitrogen vacancy
(NV)centers \cite{andrei_NV} and recently transmon qubits in the
context of circuit QED \cite{article:dispersiveQED}. In
particular, a cQED system consisting of a semiconductor
self-assembled quantum dot (QD) coupled to photonic crystal
nanocavity was used to demonstrate generation of non-classical
states of light \cite{AF_natphys,arka_tunneling,blockade_imamog},
electro-optic modulation \cite{andrei_eom}, and ultrafast all
optical switching
\cite{arka_switching,edo_switching,atac_switching}. Although, a
significant progress has been done in QD-based cQED, most of the
reported works use a neutral QD, which effectively acts as a
two-level quantum emitter with an optical frequency transition
from the ground state to the single exciton state. While such
two-level system could in principle be used as a qubit
\cite{finley_spin,waks_cnot}, the short life-time of the exciton state ($<1$
ns) makes it not suitable for practical applications.

On the other hand, the spin states of a charged QD, into which a
single electron or a hole was introduced, have been shown to
possess coherence times in the microseconds range
\cite{article:press08, article_Erik_Spin}. The use of ultrafast
optical techniques with charged QDs provides the possibility of
performing a very high number of spin manipulations within the
spin coherence time and opens avenues for their use as qubits for
quantum information applications. However, the efficiency of spin
initialization \cite{article:xu07} and manipulation achieved so
far is not high. To attain the efficiency necessary for practical
applications, one needs to enhance the light-matter interaction.
This can be achieved by embedding the charged QD in a cavity. Several groups have so far demonstrated deterministic charging
of a single QD within a photonic crystal cavity \cite{imamoglu_spin}, and magnetic field tuning of a single QD strongly
coupled to a photonic crystal cavity \cite{waks_cnot}. Recently, manipulation of a QD spin in a photonic
crystal cavity was reported by Carter et. al.
\cite{article_Gammon}. However, the configuration used in
their approach does not permit getting full advantage of the
photonic interface. We also note that a lot of effort has been directed towards spin-photon interfaces based on
NV centers in diamond \cite{lukin_entanglement}, but embedding them inside cavities has so far proven difficult.

\begin{figure}[b]
\centering
\includegraphics[width=2.8in]{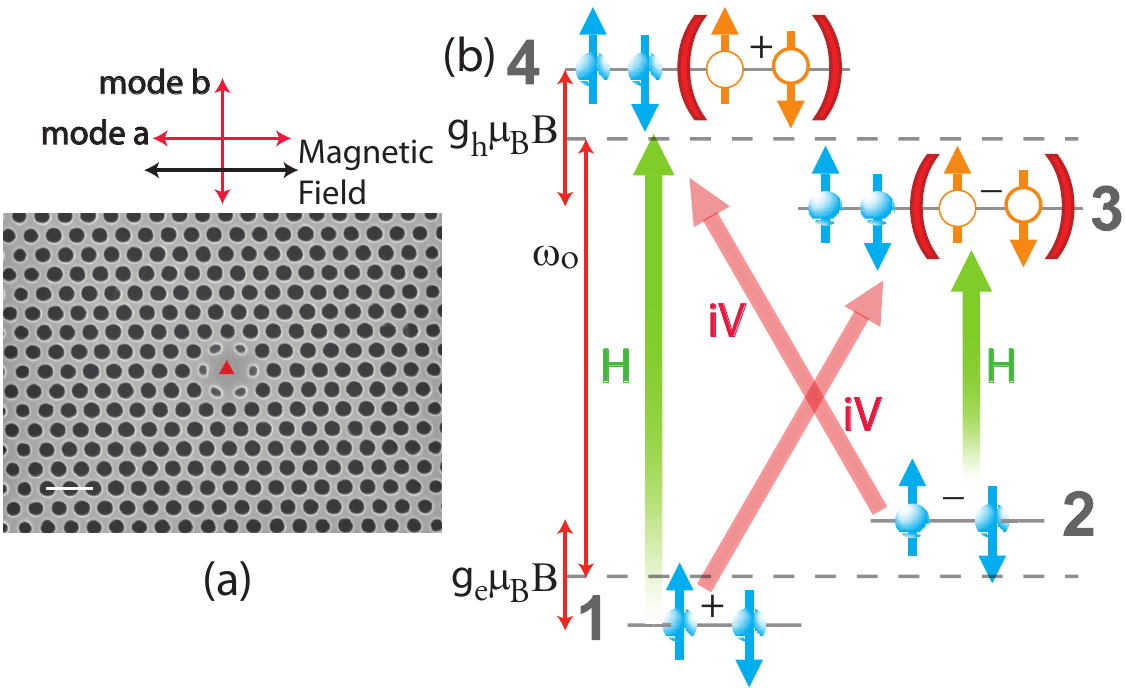}
\caption{(color online) (a) Scanning electron microscope image of
a bimodal photonic-crystal nanocavity fabricated in a GaAs
membrane with embedded quantum dots. The scalebar is $500$ nm.
(b) The schematics and the optical transitions of a four-level
system arising when a magnetic field is applied in the Voigt
configuration (see plot (a)) to a QD charged with a single
electron. The cavity mode $a$ has H polarization (see plot (a))
and can couple to transitions $|1\rangle \rightarrow |4\rangle$
and $|2\rangle \rightarrow |3\rangle$, while the cavity mode $b$
has V polarization and can drive the transitions $|1\rangle
\rightarrow |3\rangle$ and $|2\rangle \rightarrow |4\rangle$.}
\label{Spin_schematic}
\end{figure}

In this paper, we theoretically analyze a system consisting of a
QD spin coupled to a nano-cavity, with realistic system parameters. While
one might naively think that coupling a QD spin to a cavity is a
simple extension of coupling a neutral QD to a cavity, a closer
look quickly reveals that it is not so. Specifically, the QD spin
states become significantly perturbed due to presence of the
cavity, and the spin initialization or control becomes impossible when the
cavity is  brought on resonance with QD transitions, in an attempt to enhance them. In this
work, we show how this can be overcome and that
successful QD spin initialization and manipulation can be achieved
in a properly chosen configuration based on a bimodal nanocavity (Fig.
\ref{Spin_schematic}a). Bimodal photonic crystal nanocavities were previously proposed for nonclassical light generation and near-degenerate bimodal cavities were also demonstrated \cite{calzone_immamoglu,AM_bimodal}.
%Several local tuning mechanisms are also reported that can potentially be used for tuning the bimodal cavity modes in situ \cite{edo_photochromic_tuning}.
We analyze a large parameter space and find an optimal range of
detunings between QD transitions, cavity modes, and the
driving laser for which a high fidelity of spin initialization
can be achieved. Presence of a cavity also increases the speed of
spin initialization, by enhancing the rates of the coupled
transitions, bringing it to beyond GHz range, that is not achievable in bulk
semiconductor \cite{article:press08}. Finally we describe the spin manipulation in such a system. Here, we find that coherent population transfer is realized only by applying a short optical pulse that is far detuned from the QD-cavity system.

%Finally, we also show that
%in this configuration we can coherently transfer population from
%one spin state to the other by applying a short optical pulse.

%\section{Theory}
Previously, several research articles looked theoretically into
the problem of spin initialization and manipulation in a cavity
\cite{snellart_spin,AM_TP_spin,article:ima99,Rossi_spin}. However, in those studies, it was
assumed that the effect of the cavity is mainly to enhance the
local electric field of a laser and
that the QD is driven by a classical field. This
assumption breaks down for a QD embedded in a high-Q photonic
cavity and therefore quantization of the cavity field and a
careful choice of driving terms are important for a realistic
treatment of the coupled system. A single electron spin confined
to the QD can be in two different states of the same energy: spin
up or spin down, where we define the spin-state along the
optical (QD growth, i.e., $z$-axis). When a magnetic field is applied perpendicularly to
the optical axis (also known as the Voigt geometry), the spin-up
and spin-down states split by an amount $\Delta_e=g_e \mu_B B$ (Fig. \ref{Spin_schematic}b).
These states can also be thought of spin up and spin down states along the $x$-axis,
and we will use this notation for the rest of the paper.
The excited state of the QD, called the trion state, also splits.
The excited state splitting is due to the hole spin and is given by
$\Delta_h=g_h \mu_B B$. Here $g_e$
and $g_h$ are respectively, the Lande g-factor for electron and
hole; $\mu_B$ is the Bohr magnetron; $B$ is the applied magnetic
field; and in this work we neglect any diamagnetic shift of the
QD. The lossless dynamics of the system consisting of QD spins
coupled to a cavity with two modes of perpendicular polarizations
(labeled  $H$ and $V$ in Fig. \ref{Spin_schematic}b) and driven by
a laser, can be described by the Hamiltonian
\begin{equation}
\label{eqn:H_res_0}
\mathcal{H}=\mathcal{H}_o+\mathcal{H}_{int}+\mathcal{H}_d
\end{equation}
where in the rotating-wave approximation (and with $\hbar=1$)
\begin{eqnarray}
\nonumber
\mathcal{H}_o&=& -\frac{\Delta_e}{2}|1\rangle\langle1|+\frac{\Delta_e}{2}|2\rangle\langle2|
+(\omega_o-\frac{\Delta_h}{2})|3\rangle\langle3|\\
&&+(\omega_o+\frac{\Delta_h}{2})|4\rangle\langle4|+\omega_{a} a^\dag a + \omega_{b} b^\dag b\\
\mathcal{H}_{int}&=&g_aa^\dag (\sigma_{14}+\sigma_{23})+ig_bb^\dag (\sigma_{24}+\sigma_{13})+h.c.
\end{eqnarray}
Here, $g_a$ and $g_b$ describe the coupling strengths between the
cavity modes and the QD transitions, $a$ and $b$ are the photon
annihilation operators of the two cavity modes with frequencies
$\omega_a$ and $\omega_b$, respectively, $\sigma_{ij}=\vert
i\rangle \langle j\vert$, and $\omega_o$ is the frequency of the
QD's optical transitions in the absence of the magnetic field (see
Fig. \ref{Spin_schematic} b).  The driving part of the Hamiltonian,
$\mathcal{H}_d$, changes depending on whether the applied laser
field drives the QD or the cavity.
%In an experiment we always drive both the QD and the cavity. However, when the laser is tuned to a resonant QD-cavity system, then mostly the cavity is driven. %For a detuned system we assumed that mostly the QD is driven.
We consider the general form of
the driving Hamiltonian $(\mathcal{H}_d=\mathcal{H}_d^{cav}+\mathcal{H}_{d}^{QD})$, which describes a single laser with
controllable polarization capable of driving both the cavity
modes ($\mathcal{H}_d^{cav}$) or the QD directly ($\mathcal{H}_d^{QD}$):
\begin{equation}
\mathcal{H}_d^{cav}=\mathcal{E}_ae^{i\omega_l t}a + \mathcal{E}_be^{i\omega_l t}b+h.c.
\end{equation}
%where for simplicity we assume that the applied laser field excites both cavity modes equally.
where $\mathcal{E}_{a,b}$ are the rates with which the applied
laser field excites each of the cavity modes, and
\begin{equation}
\mathcal{H}_d^{QD}=\Omega_h e^{i\omega_l t}(\sigma_{13}+\sigma_{24}) + \Omega_v e^{i\omega_l t}(\sigma_{23}+\sigma_{14})+h.c.
\end{equation}
Here, $\Omega_h$ and $\Omega_v$ are the Rabi frequencies of the laser for the horizontal and the vertical QD transitions, respectively.
Depending on the driving conditions, one of the two terms in $\mathcal{H}_d$ can be dominant. For example, if a laser can couple to cavity resonance, we assume that $\mathcal{H}_d^{cav}$ is dominant, and  neglect $\mathcal{H}_d^{QD}$. In other words, QD is always driven via a cavity mode. This condition is assumed for spin initialization. On the other hand, if the laser cannot couple well to cavity resonance, but QD transitions are instead driven directly, $\mathcal{H}_d^{QD}$ dominates. This happens when the laser detuning from QD transitions is smaller than from cavity resonances \cite{majumdar_QD_splitting}, or when the laser is applied from the spatial direction where it does not couple to cavity modes. In this paper, we show that for coherent spin manipulation, it is necessary to drive the QD directly, and not via cavity mode. We can transform the Hamiltonian into the rotating frame by using
$\mathcal{H}_{r}=T^\dag\mathcal{H}T+i\frac{\partial
T^\dag}{\partial t}T$ where $T=e^{-i\omega_lt(a^\dag
a+b^\dag b+|3\rangle\langle3|+|4\rangle\langle4|)}$.

%to find that
%$\mathcal{H}_{r}=\mathcal{H}_{ro}+\mathcal{H}_{rint}+\mathcal{H}_{rd}$
%with
%\begin{eqnarray*}
%\mathcal{H}_{ro}&=& -\frac{\Delta_e}{2}|1\rangle\langle1|+\frac{\Delta_e}{2}|2\rangle\langle2|
%+(\Delta_o-\frac{\Delta_h}{2})|3\rangle\langle3|\\
%&&+(\Delta_o+\frac{\Delta_h}{2})|4\rangle\langle4|+\Delta_{c1} a^\dag a+ \Delta_{c2}  b^\dag b\\
%\mathcal{H}_{rint}&=&g_1a^\dag (\sigma_{14}+\sigma_{23})+g_2b^\dag (\sigma_{24}+\sigma_{13})+h.c.\\
%\mathcal{H}_{rd}&=&\mathcal{E}_aa+\mathcal{E}_bb+h.c.
%\end{eqnarray*}
%where, $\Delta_o=\omega_o-\omega_l$ and $\Delta_{c1,2}=\omega_{c1,2}-\omega_l$.

The losses in the system are incorporated by solving the master equation of the density matrix $\rho$ of the coupled QD-cavity system:
$\frac{d\rho}{dt}=-i[\mathcal{H}_{r},\rho]+\sum_j \mathcal{L}(c_j)$
where, $\mathcal{L}(c_j)=2c_j \rho c_j^{\dag}-c_j^{\dag} c_j \rho-\rho c_j^{\dag} c_j$ is the Lindblad operators for the
collapse operator $c_j$. In this case, we have six different loss
channels, and hence six collapse operators (one for each cavity
and each QD transition): $c_j\in \{ \sqrt{\kappa} a, \sqrt{\kappa} b,
\sqrt{\gamma_{41}}\sigma_{41},
\sqrt{\gamma_{42}}\sigma_{42},\sqrt{\gamma_{31}}\sigma_{31},
\sqrt{\gamma_{32}}\sigma_{32}\}$.

%In our numerical simulations of the
%master equation, we truncate the basis of the cavity modes to
%$N=4$ Fock states. We note that, we use such a small basis for the
%photons to maintain a tractable Hilbert space size (in this case
%$100 \times 100$). However, we verified our findings also by using
%stochastic Schr$\ddot{o}$dinger equations \cite{arka_tunneling}, where we
%consider only the wave-function of the coupled system, and the
%computation is easier. We find that although a bigger Hilbert
%space is required for predicting the cavity observables,
%relatively little dependence on the size of Hilbert space is
%noticed for the QD observables.

\begin{figure}
\centering
\includegraphics[width=3.5in]{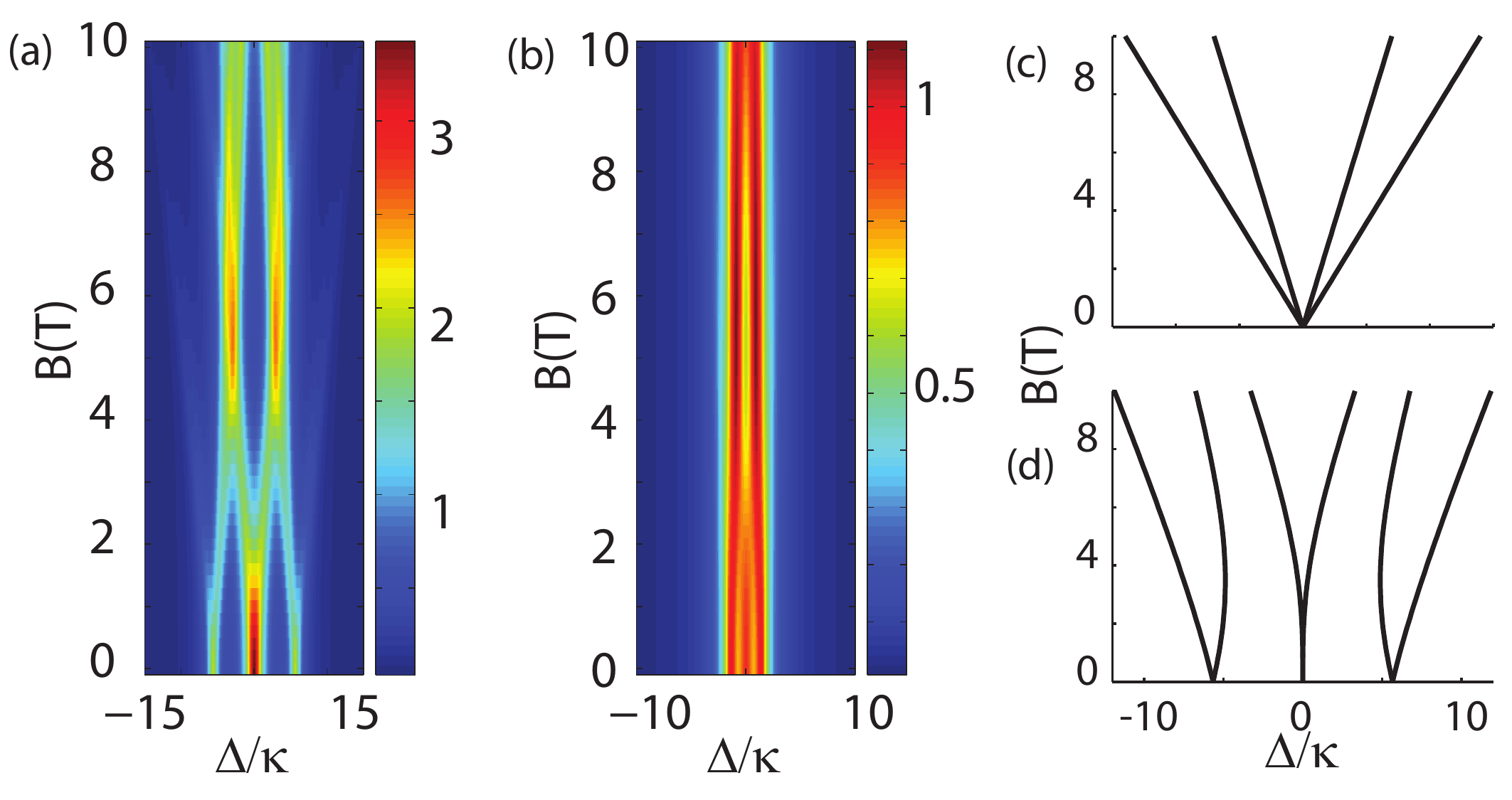}
\caption{(color online) Photoluminescence spectra of the coupled
QD spin-cavity system as a function of the magnetic field: (a)
for a QD-cavity system with: $\kappa/2\pi=5$ GHz and $g/2\pi=20$
GHz.(b): similar PL spectra for  $g/2\pi=\kappa/2\pi=20$
GHz. (c),(d) Energy splitting between all the QD transitions as a
function of the magnetic field without(c) and with (d) a cavity,
in the absence of losses ($\kappa=0$).} \label{Fig_PL_charac}
\end{figure}

%\begin{figure}
%\centering
%\includegraphics[width=3.5in]{Fig_refl_charac.pdf}
%\caption{(color online) Transmission spectra of the coupled QD spin-cavity system as a function of the increasing magnetic field: (a),(b),(c) for system with $\kappa/2\pi=5$ GHz and $g/2\pi=20$ GHz and (d),(e),(f) for system with $g/2\pi=\kappa/2\pi=20$ GHz. For (a),(d) we drive and collect from the mode $a$; for  (b),(e) we drive and collect from the mode $b$ and for (c), (f) we drive both $a$ and (b) and collect from the both modes.}
%\label{Fig_refl_charac}
%\end{figure}

%\section{Coupled Charged QD-Cavity System}
We now use this model to theoretically investigate the spectrum of
the coupled charged QD-cavity system probed under
photoluminescence (PL) as is commonly done experimentally (to
reveal eigenstates of the system). We perform this analysis for
two different cavity decay rates: readily achievable cavity decay
rate: $\kappa/2\pi=20$GHz and better than the state of the art
(but achievable with cavity fabrication improvements) system
parameters: $\kappa/2\pi=5$ GHz. The dot-cavity interaction
strength is $g/2\pi=20$ GHz for both cases (corresponding to
experimentally achievable condition \cite{article:majumdar09}). The dipole decay rates are $\gamma_{41},
\gamma_{42},\gamma_{31},
\gamma_{32}/2 \pi=1$ GHz. When the system is characterized through PL, an above
band laser pumps the semiconductor to generate electron hole
pairs. These carriers recombine in the QD, which subsequently emits a photon. This is an
incoherent way of probing the system, and is modeled by adding
Lindblad terms $P(\mathcal{L}(a^\dag)+\mathcal{L}(b^\dag))$, which
signify incoherently populating the cavities with a rate $P$. A low value of $P/2\pi \sim 0.1$ is used to allow using a small Fock state basis $(N=4)$.
We calculate the power spectral density (PSD) $S(\omega)$ of the
system given by
\begin{equation}
S(\omega)=\int_\infty^\infty <a^\dag(t) a>e^{-i\omega t}+<b^\dag(t) b>e^{-i\omega t} dt
\end{equation}
where $\omega$ is the spectrometer frequency, and in rotating frame $\Delta=\omega-\omega_o$.
The numerically simulated PSD as a function of increasing magnetic
field is shown in Fig. \ref{Fig_PL_charac} for two different
values of the cavity decay rates $\kappa/2\pi=5$ GHz and
$\kappa/2\pi=20$ GHz. Peaks in
the PSD correspond to eigen-states of the coupled system. For a system with low $\kappa$, we observe six peaks
(for $B>0$). However, with increasing cavity decay rates, the
energies of the eigenstates become degenerate and such structure
disappears. To understand the origin of the six peaks, we note
that with one quantum of energy present in the system, the bare
states of the coupled charged QD-bimodal cavity system are
$|1,0,1>$, $|0,1,1>$, $|1,0,2>$, $|0,1,2>$, $|0,0,3>$ and
$|0,0,4>$, where the first number denotes the number of photons
present in the mode $a$, second number is the number of photons in
mode $b$ and the last one is the populated charged QD state. If
$g/\kappa$ is sufficiently large, these bare states couple and
give rise to six dressed states that we observe in the PL spectrum
in Fig. \ref{Fig_PL_charac} (a). For additional intuitive
understanding, we diagonalize the Hamiltonian when only one photon
is present in the system and plot the eigen-values as a function
of the magnetic field. In absence of the cavity, we see four
transitions from the QD (Fig. \ref{Fig_PL_charac}c). In presence
of the two cavity modes, however, a hybridization between the
cavity modes and the QD transitions occurs, which results in six
observable transitions (Fig. \ref{Fig_PL_charac}d). We note that
these six transitions are also present in the system with larger
cavity losses (Fig. \ref{Fig_PL_charac} b), but they cannot be
as clearly resolved as in Fig. \ref{Fig_PL_charac} a because of
their overlap.

%\section{Spin Initialization}
\begin{figure}
\centering
\includegraphics[width=3.5in]{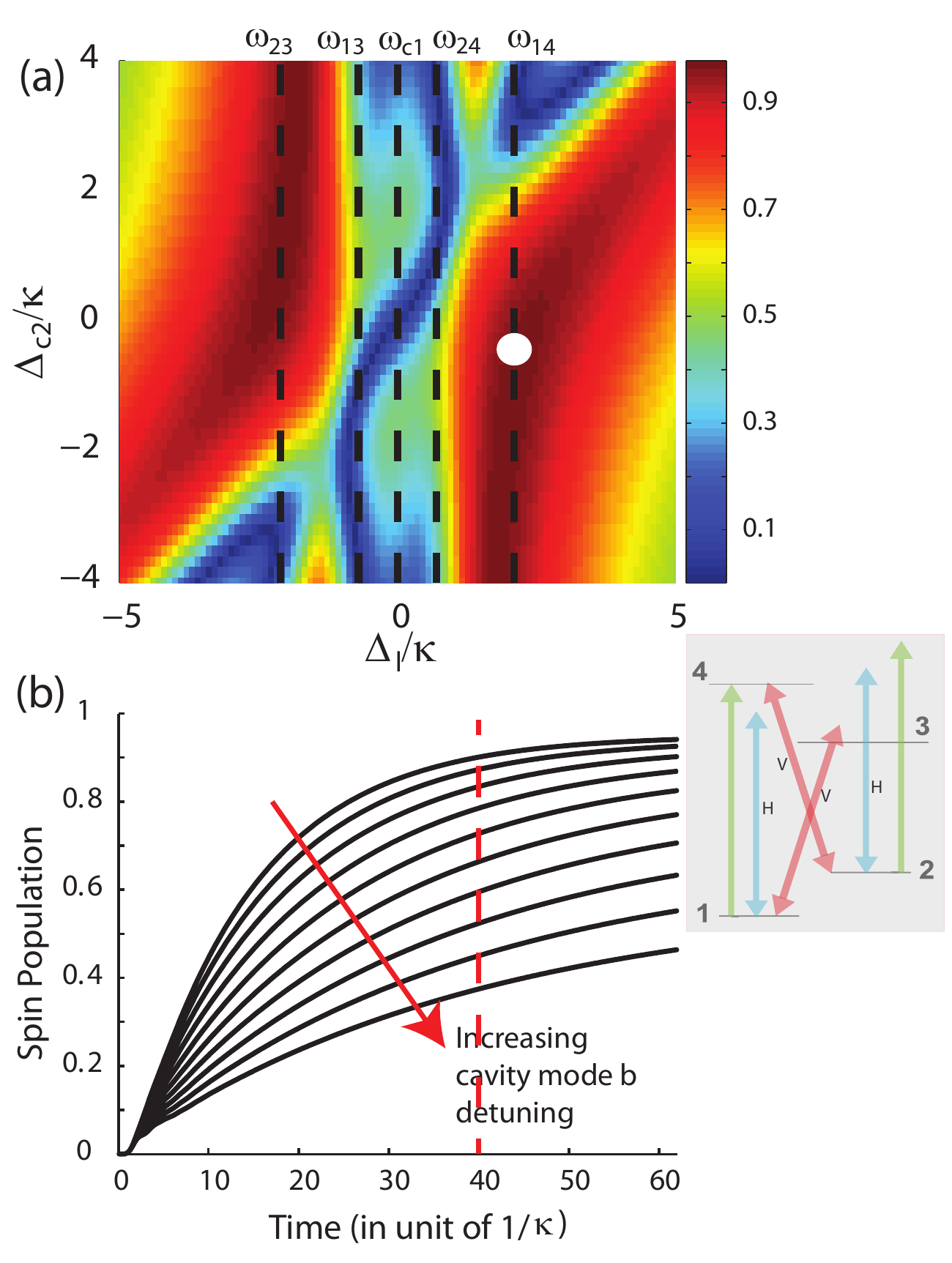}
\caption{(color online) The fidelity and speed of spin
initialization for a realistic QD-cavity system
($g/2\pi=\kappa/2\pi=20$ GHz at a magnetic field of $5$ T). (a)
Initialization fidelity $\vert \rho_{11}-\rho_{22} \vert$ as a
function of the cavity mode $b$ frequency
$\Delta_{b}=\omega_{b}-\omega_o$ and the pump laser wavelength
$\Delta_l=\omega_l-\omega_o$. The white point marks the situation
where we get optimal spin initialization fidelity (the situation
shown in the inset of (b): the pump laser is tuned to the QD transition $\vert 1\rangle
\rightarrow \vert 4 \rangle$, and the cavity mode $b$ is tuned to
the transition $\vert 2 \rangle \rightarrow \vert 4 \rangle$. Here double arrows denote cavity fields and the single arrows
denote the laser). (b) The spin initialization as a function of the
time for different cavity detunings $\Delta_{c2}$, with laser
being fixed at the transition frequency $\omega_{14} (\Delta_l
\sim 2\kappa)$. $\Delta_{c2}$ is changed from $0$ to $2\kappa$. We
note that the spin-initialization time is only around $40/\kappa
\sim 300$ ps.} \label{Figure_init_detuning}
\end{figure}

Next, we proceed to study the spin initialization
in such a QD-cavity system. We analyze the speed and fidelity of
spin initialization as a function of laser and cavity detuning. We
consider the state of the art parameter set for these simulations
($\kappa/2\pi=20$ GHz and $g/2\pi=20$ GHz) and assume a magnetic
field of $5$ T, resulting in $\Delta_e/2\pi \approx 28$ GHz and
$\Delta_h/2\pi \approx 14$ GHz. We also assume that the QD spin
starts in a mixed state with equal spin up and down (i.e., equal states $|1\rangle$ and $|2\rangle$)
population: $\rho_{11}=\rho_{22}=1/2$. We pump the
system with an H polarized laser (which couples to mode $a$),
while the other (V-polarized) cavity mode $b$ is not driven. In
other words, we use the laser to drive outer (H) transitions in Fig.
\ref{Spin_schematic}b via cavity mode $a$, but the inner (V)
polarized QD transitions are coupled to vacuum field of the cavity
mode $b$. For each $\Lambda$ QD system, this resembles the vacuum
induced transparency (VIT) configuration recently proposed and
experimentally studied in atomic physics
\cite{vuletic_vit}. Fig. \ref{Figure_init_detuning}(a) plots
$\vert \rho_{11}-\rho_{22} \vert$ as a function of the pump laser
wavelength and the cavity mode $b$ frequency $\omega_{b}$. The
cavity mode $a$ frequency $\omega_{a}$ is kept fixed at
$\omega_o$ (QD transition in absence of $B$-field, see Fig.\ref{Spin_schematic}), so the laser is not
necessarily on resonance with this mode. We find that a high
fidelity spin initialization is achieved when the pump laser is
tuned to the QD transition $\vert 1\rangle \rightarrow \vert 4
\rangle$, and the cavity mode $b$ (which is in vacuum state) is
tuned to the transition $\vert 2 \rangle \rightarrow \vert 4
\rangle$ (see inset of Fig. \ref{Figure_init_detuning}(b)). The laser can
potentially couple to the transition $\vert 2 \rangle \rightarrow
\vert 3\rangle$ and the cavity mode $b$ to the transition $\vert
1\rangle \rightarrow \vert 3\rangle$. However, due to detunings of
the laser and the cavity mode $b$, the QD is efficiently optically
pumped only via $\vert 1 \rangle \rightarrow \vert 4
\rangle\rightarrow \vert 2\rangle$ route, leading to all the spin
population in the QD state $\vert 2 \rangle$. In a similar fashion
we can also use the path $\vert 2 \rangle \rightarrow \vert
3\rangle \rightarrow \vert 1\rangle$, with a different set of
detunings to achieve initialization in state $\vert 1\rangle$.
Fig.\ref{Figure_init_detuning}a also reveals that a high spin
initialization fidelity can also be achieved when the cavity is
far detuned \cite{article_Gammon}, but this is a trivial case equivalent to the
situation of a QD uncoupled to the cavity  (a QD in bulk
semiconductor), eliminating the cavity's beneficial role.

Next, we focus on the temporal dynamics of the spin-population
with varying detuning $\omega_{b}$ of the cavity mode $b$. We
observe that the spin-initialization is faster when the cavity is
resonant to the $|2\rangle \rightarrow |4\rangle$ transition
(Fig.\ref{Figure_init_detuning}(b)). The initialization speed of
several GHz is achieved in this case, but even faster
initialization speed can be achieved by pumping the system with a
stronger laser (while keeping a resonant cavity on the other
transition in the $\Lambda$ system). The simulations were performed using the open source python package QuTip \cite{qutip}.

\begin{figure}
\centering
\includegraphics[width=3.25in]{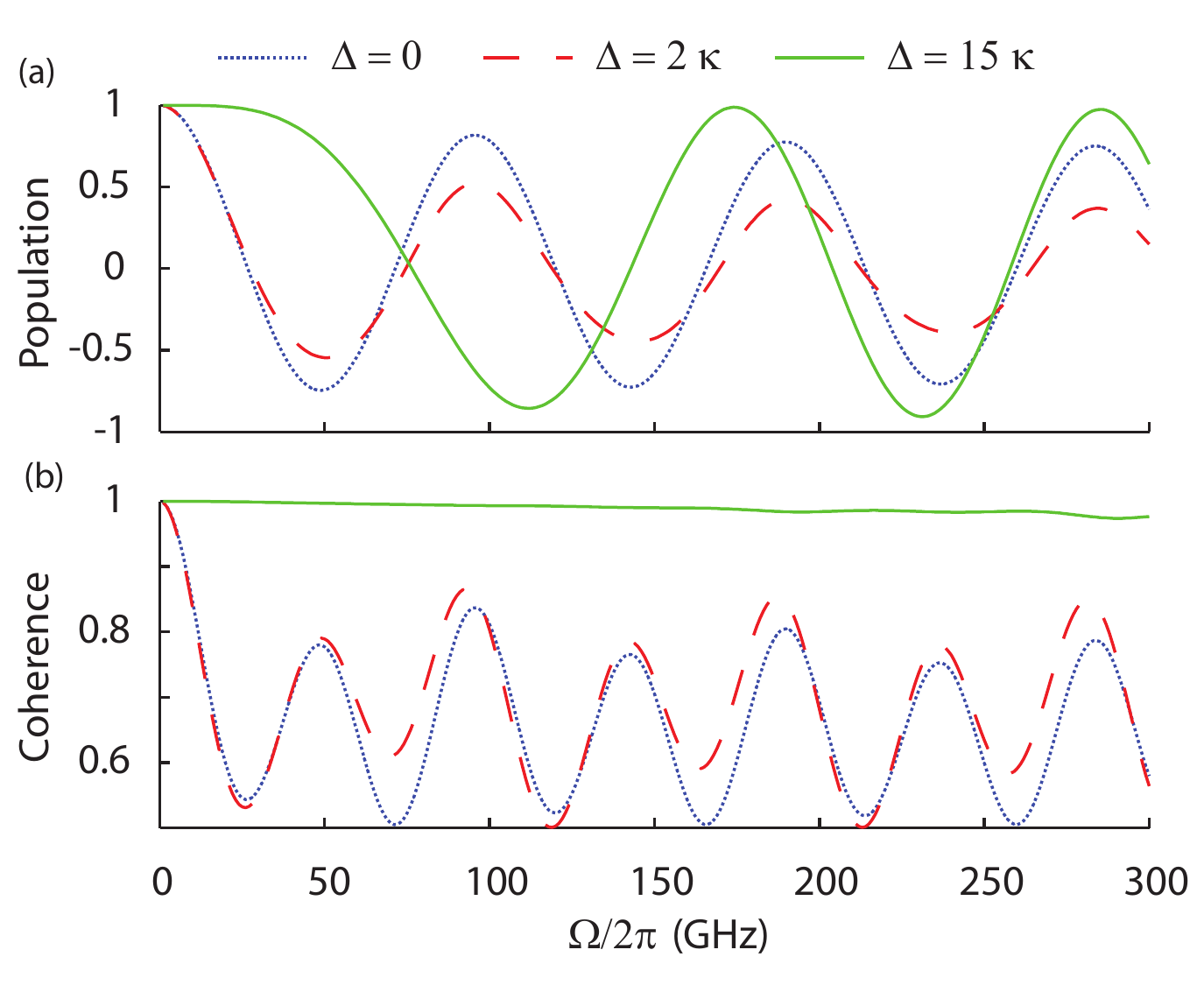}
\caption{(color online) (a) The steady-state spin population ($\rho_{11}-\rho_{22}$) as a function of
the laser Rabi frequency ($\Omega_h=\Omega_v$). All population is initially in state $|1\rangle$, and a $5$
ps laser pulse excites the system. Three different pulse detunings are used. Rabi oscillations between two spin states are
observed. (b) $Tr[(\rho^{(1,2)})^2]$ as a function of the laser Rabi frequency. The process becomes coherent ($Tr[(\rho^{(1,2)})^2]\sim 1$) when the detuning of the pulse from the QD-cavity system is large. }\label{Figure_spin_manipulate}
\end{figure}

%\section{Spin Manipulation}
Finally we analyze how the spin-manipulation can be performed in
such a system. The coupled QD-cavity system is driven by a laser pulse with different detunings,
and the spin-population is monitored as a function of the pulse
amplitude. Initially all the population is in the state $|1\rangle$.
We assume that the cavity is at the undressed QD frequency $\omega_o$, and we drive the QD directly
(i.e., the driving conditions are such that the laser is spatially or spectrally decoupled from cavity modes) \cite{supplementary}. The system is excited with a short pulse (pulse width $5$ ps), and the spin
population in states $|1\rangle$ ($\rho_{11}$) and $|2\rangle$ ($\rho_{22}$) are monitored over time \cite{supplementary}.
The spin population difference ($\rho_{11}-\rho_{22}$) in the steady state is plotted as a function of the pulse amplitude for three different pulse detunings $\Delta$ from the cavity resonances (Fig. \ref{Figure_spin_manipulate} a). Rabi oscillations are observed between the spin up ($|1\rangle$) and spin
down ($|2\rangle$) states as the pulse amplitude is changed.
%Note that for these simulations we used a larger Fock state basis $(N=12)$.
To check whether the process is coherent, we calculate the trace of the density matrix $\rho^{(1,2)}$ of the subspace consisting of spin up and down states, and plot $Tr[(\rho^{(1,2)})^2]$ in Fig. \ref{Figure_spin_manipulate} b. $Tr[(\rho^{(1,2)})^2]$ is unity for a pure state. Therefore, if the spin manipulation process is coherent, we expect this value to be near unity. Our results in Fig. \ref{Figure_spin_manipulate}b indicate that only at a large detuning $\Delta$, the process is coherent. Additionally, the process is only coherent if the QD is driven directly (and not via cavity) \cite{supplementary}. Therefore, a coherent spin manipulation in the proposed system is also possible, but only by applying an optical pulse that is far detuned from the cavity and drives the QD directly.

%\section{Conclusion}
In summary, we have presented a proposal for efficient
initialization and manipulation of a single QD spin coupled to a
bimodal optical cavity. Our numerical analysis (with full field
quantization and with realistic system parameters) confirms that
nearly $100$\% spin initialization fidelity is achievable with
speed beyond GHz, as well as spin manipulation, benefiting from a
cavity not only as a photonic interface but also to speed up the
spin control.

%\section{Acknowledgement}
The authors acknowledge financial support provided by the Air Force Office of Scientific Research, MURI Center for Multi-functional light-matter interfaces based on atoms and solids. K.G.L. acknowledges support by the Swiss National Science Foundation and A.R. is also supported by a Stanford Graduate Fellowship.

\appendix
\section{Hamiltonian}
In this section, we analyze the terms in the Hamiltonian of the coupled QD-cavity system. As shown in Fig. $1$ (in the main paper), the horizontal and vertical transitions have a $\pi/2$ phase difference, so we assume that the horizontal dipole moment of the QD is $\mu_H$ and the vertical dipole moment is $i\mu_V$. On the other hand, the cavity has two polarizations $H$ and $V$, and the QD-cavity interaction strength for the horizontal cavity is given by $g_a$, and for the vertical cavity it is given by $i g_b$.

\begin{figure}
\centering
\includegraphics[width=3.25in]{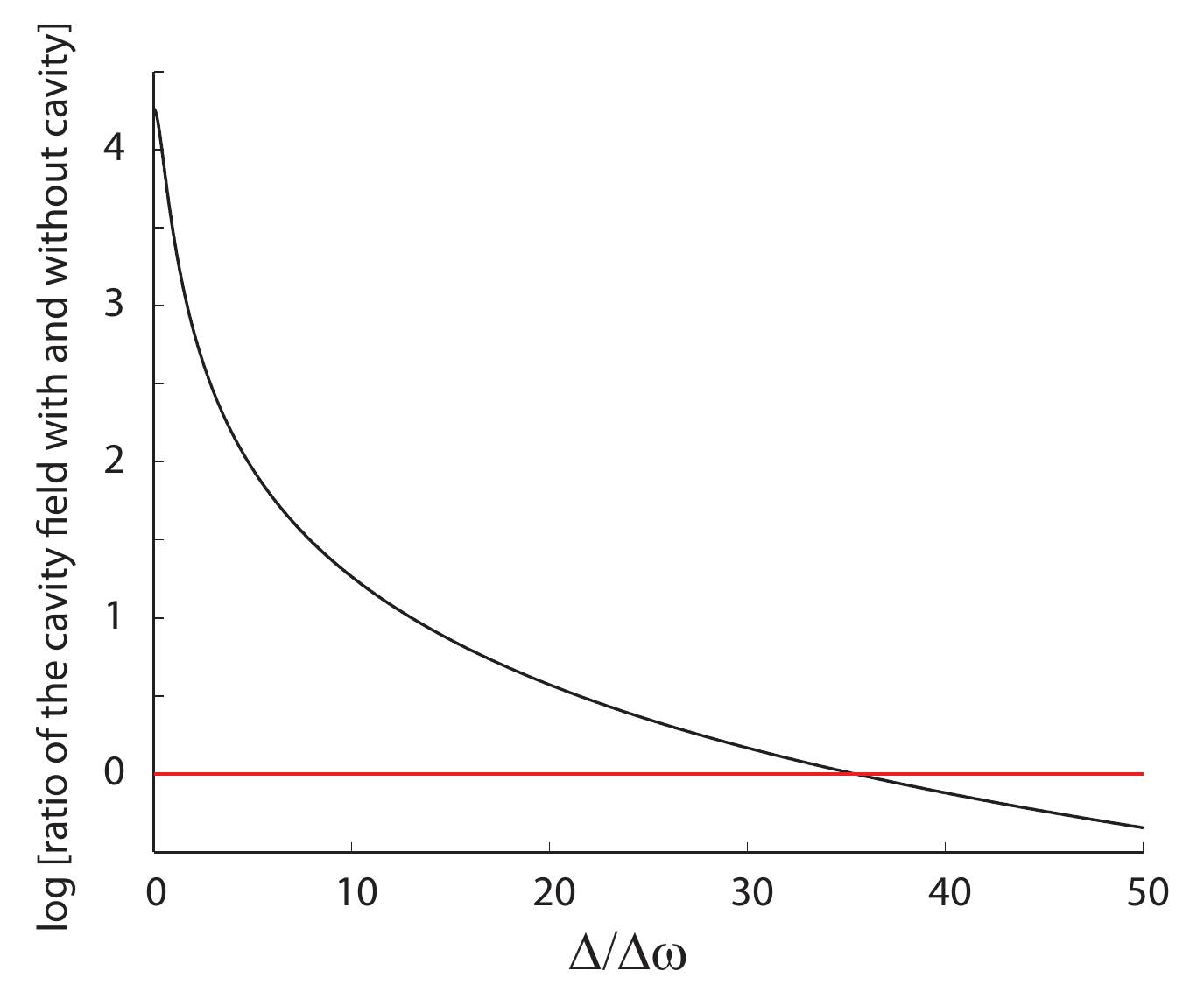}
\caption{The ratio of electric field for the driving laser for a QD without any cavity, and the electric field inside the cavity, as a function of the detuning between the laser and the cavity resonance. The QD is assumed to be at near resonance to the cavity. For the simulation we assumed: $\eta=0.1$; $V_m=(\lambda/n)^3$; $\kappa/2\pi=20$ GHz. The red line is when $E_{cav}=E_{nocav}$.} \label{fig_ratio_cav_nocav}
\end{figure}

As described in the main paper, depending on the driving condition, one of the two terms in the driving Hamiltonian dominates. The electric field of the laser driving the QD directly (field in the absence of any cavity, or laser doesn't couple spatially or spectrally to cavity mode) is given by \cite{majumdar_QD_splitting}:
\begin{equation}
|E_{nocav}|=\frac{2}{1+n}\sqrt{\frac{P}{c\epsilon\pi \sigma_0^2}}
\end{equation}
where $n$ is the refractive index of GaAs; $P$ is the power of the Gaussian laser beam; $c$ is the velocity of light; and $\sigma_0$ is the radius of the Gaussian beam. On the other hand the electric field of the laser that builds up in the cavity is given by
\begin{equation}
\label{eqn_power_field} |E_{cav}|=\sqrt{\frac{\eta P Q
\lambda_0}{2\pi c \epsilon
V_m}\frac{1}{1+(2\Delta/\Delta\omega)^2}}
\end{equation}
where, $Q$ is the quality factor, $V_m$ is the mode volume of the cavity, $\Delta\omega$ is the linewidth of the cavity, and $\Delta$ is the detuning of the the laser from the cavity. The ratio between the two electric fields is given by
\begin{equation}
\frac{E_{cav}}{E_{nocav}}=\frac{1+n}{2}\sqrt{\frac{\eta Q
\lambda_0
\sigma_0^2}{2V_m}\frac{1}{1+(2\Delta/\Delta\omega)^2}}
\end{equation}
Fig. S.\ref{fig_ratio_cav_nocav} shows the ratio as a function of $\Delta$. As expected, we find that when the laser is near-resonant to the cavity, then the electric field inside the cavity is much larger than the electric field without the cavity. In that case, we can assume that $\mathcal{H}_d^{cav}$ is dominant. We note that, when the laser is near-resonant to the cavity, it is not accurate to assume that all the enhanced electric field is driving the QD, as the QD may get saturated. On the other hand, if the laser is spectrally far detuned from the cavity resonance or spatially decoupled from it, the situation is different, and the direct QD driving term ($\mathcal{H}_d^{QD}$) dominates. For example, if the laser is resonant to the QD, and QD is off-resonant from the cavity, then a QD driving is appropriate \cite{majumdar_QD_splitting}. Similarly, for a largely detuned pulse from the cavity, a QD driving is appropriate, no matter where the QD is spectrally located, as a far-detuned pulse from cavity is similar to a pulse driving a QD in bulk. On the other hand, if a QD in a micropost is driven from the side \cite{article:michler09}, then the laser is spatially decoupled from the cavity, and a QD driving is appropriate. In the paper, we assume that driving conditions for spin initialization are such that $\mathcal{H}_d^{cav}$ dominates, and for manipulation that $\mathcal{H}_d^{QD}$ dominates. This is valid, as for initialization we assumed that the continuous wave (cw) pump laser is near-resonant to the cavity (also near-resonant to the QD). On the other hand, for coherent spin manipulation, we are driving a near-resonant QD-cavity system with a very far detuned pulse $(\Delta \sim 15\kappa)$. It should be noted that, there are situations when both driving terms should be kept, for example, if we drive with a pulse which is slightly detuned from a near-resonant dot-cavity system. In the main paper, we show that such situations are detrimental for coherent spin control irrespective of whether we include the $\mathcal{H}_d^{cav}$ term or not.

The Hamiltonian in the rotating frame describing the QD driving is given by $\Omega_{v} (\sigma_{23}+\sigma_{14})+\Omega_{h} (\sigma_{13}+\sigma_{24})+h.c.$, where $\sigma_{ij}$ and $\sigma_{ij}^\dag$ are the destruction and the creation operators for the QD transitions. Here $\Omega$ is the Rabi frequency of the laser and is given by
\begin{equation}
\Omega=\frac{\mu.E}{\hbar}
\end{equation}
where $E$ is the electric field of the driving laser. If we drive the QD spin transitions (the horizontal dipole moment is $\mu_H$ and the vertical dipole moment is $i\mu_V$) with circularly polarized light $S_{\pm}=H\pm iV$, then the Rabi frequencies in the driving term are real as stated above. However, if we drive the QD spin transitions with a diagonally ($45^o$) polarized light, then there is a $\pi/2$ phase difference in the Rabi frequencies, i.e. the Hamiltonian becomes $i\Omega_{v} (\sigma_{23}+\sigma_{14})+\Omega_{h} (\sigma_{13}+\sigma_{24})+h.c.$. Note that, with such $45^o$ pulses we cannot perform spin rotation. For driving a cavity however, the driving term is given by $\mathcal{E}_a a+\mathcal{E}_b b+h.c.$, where $a,b$ and $a^\dag,b^\dag$ are the annihilation and creation operators for the cavity photons. The driving strengths $\mathcal {E}$ are proportional to $\sqrt{\kappa}$, $\kappa$ being the cavity field decay rate, and does not depend on the dipole moment of the QD. Hence for a circularly polarized light, the driving term will have a phase difference between the horizontal and the vertical polarization, i.e. $\mathcal{E}_a a+i\mathcal{E}_b b+h.c.$.

\section{Effect of cavity driving on coherence}
\begin{figure}
\centering
\includegraphics[width=3.25in]{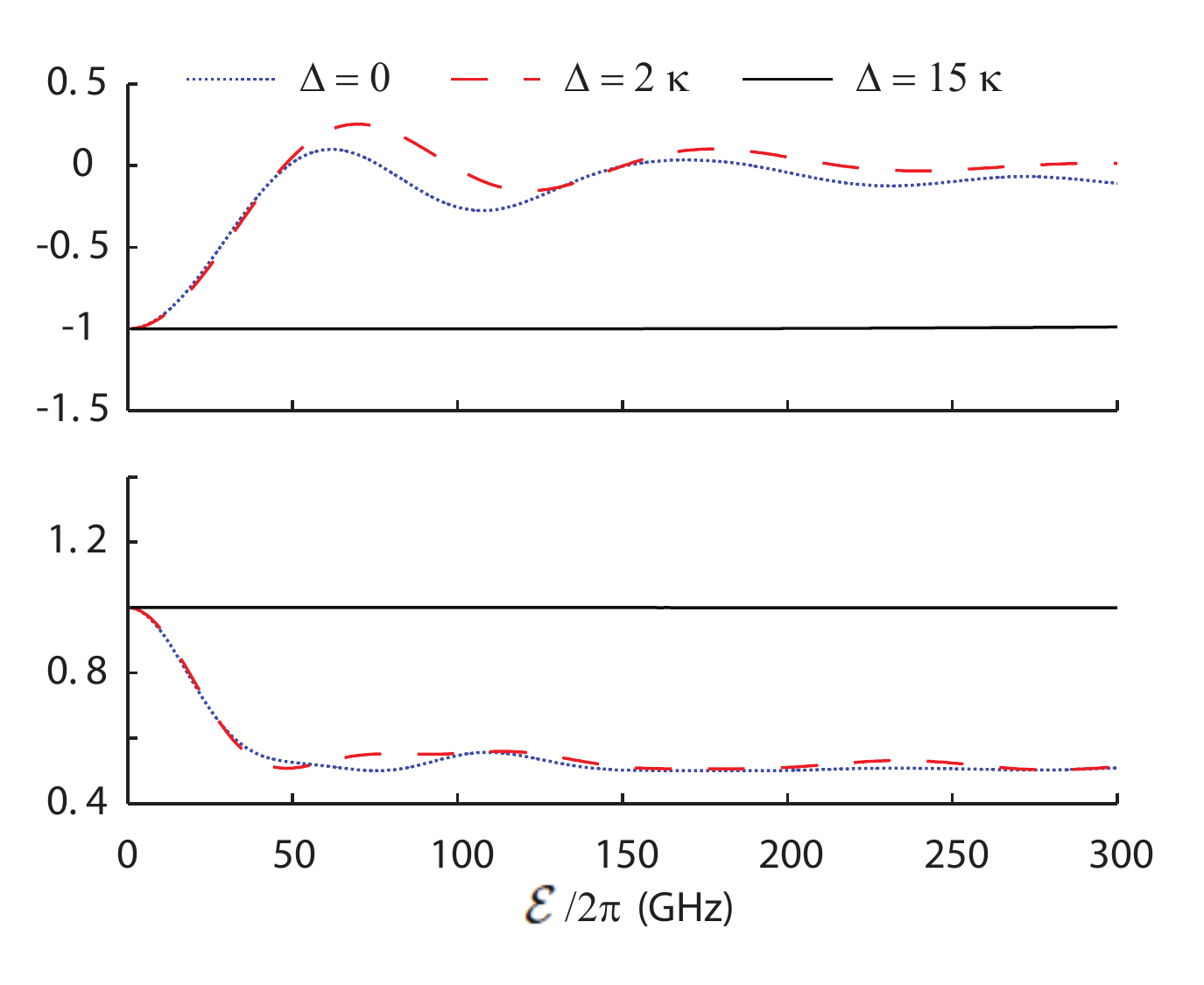}
\caption{(color online) (a) The steady-state spin population ($\rho_{11}-\rho_{22}$) as a function of
the cavity driving rate $\mathcal{E}=\mathcal{E}_a=\mathcal{E}_b$. All population is initially in state $|1\rangle$, and a $5$
ps laser pulse excites the system. Three different pulse detunings are used. (b) $Tr[(\rho^{(1,2)})^2]$ as a function of the laser Rabi frequency. }\label{cavity_driving_spin}
\end{figure}

In the main paper, we showed that a coherent spin control is possible, only when the coupled QD-cavity system is driven by a far detuned pulse, and we modeled this process only by having $\mathcal{H}_d^{QD}$ and ignoring $\mathcal{H}_d^{cav}$. Although, such an assumption is valid for a far detuned pulse from the cavity ($\Delta \sim 15 \kappa$, where we get back a coherent process, as shown in the Fig. $4$b of main paper), the assumption fails when the detuning $\Delta$ reduces. At a reduced detuning, the cavity is also driven, as the pulse is no longer spectrally decoupled from the cavity. We also analyze such cavity driving (Fig. S. \ref{cavity_driving_spin}), for the case of a circularly polarized driving laser. We note that for small laser-cavity detuning, the cavity is strongly driven and many photons are excited, hence a relatively large photons basis must be employed to achieve convergence. We find that for a far detuned pulse ($\Delta \sim 15\kappa$), and cavity driving, we cannot do any spin manipulation, regardless of whether circularly polarized light is employed or not. The population remains in the state $|1\rangle$ of the QD only. This supports our assumption that at such large detuning the driving of the cavity is negligible, and driving the QD is the correct model. At smaller detunings, however oscillations in the population are observed. But at these detunings the process is no longer coherent. Thus we find that under the cavity driving, we can never perform coherent spin manipulation, and direct QD driving is required.

\bibliography{Spin_Photon_PR}
\end{document}